 \documentclass[aps,pre,reprint,groupedaddress]{revtex4-1} 

\usepackage{graphicx}
\usepackage{amsmath}
\usepackage{amssymb}
\usepackage{bm}
\usepackage{cancel}
\usepackage[normalem]{ulem}
\usepackage{xfrac}
\usepackage[colorlinks=true,linkcolor=blue,citecolor=blue,urlcolor=blue]{hyperref}

\usepackage{textcomp}
\usepackage{color}
\usepackage{xcolor}

\usepackage{hyperref}
\usepackage{cleveref}

\usepackage{subfigure}
\usepackage{placeins}
\usepackage{siunitx}
\usepackage{blindtext}

\begin{document}

\title{Tuning and jamming reduced to their minima }

\author{Miguel Ruiz-Garc\'ia}
\affiliation{Department of Physics and Astronomy, University of Pennsylvania, Philadelphia, PA 19104, USA}

\author{Andrea J. Liu}
\affiliation{Department of Physics and Astronomy, University of Pennsylvania, Philadelphia, PA 19104, USA}

\author{Eleni Katifori}
\affiliation{Department of Physics and Astronomy, University of Pennsylvania, Philadelphia, PA 19104, USA}

\date{\today}

\begin{abstract}
Inspired by protein folding, we smooth out
the complex cost function landscapes of two processes, the tuning of networks, and the jamming of ideal spheres. In both processes, geometrical frustration plays a role -- tuning pressure differences between pairs of target nodes far from the source in a flow network impedes tuning of nearby pairs more than the reverse process, while unjamming the system in one region can make it more difficult to unjam elsewhere. By modifying the cost functions to control the order in which functions are tuned or regions unjam, we smooth out local minima while leaving global minima unaffected, increasing the success rate for reaching global minima.
\end{abstract}
\maketitle

\section{Introduction}

Many problems from physics to computer science involve the minimization of some free energy, loss or cost function in a high-dimensional space defined by the degrees of freedom (phase space). Such problems generally fall into the class of constraint-satisfaction problems, with cost functions given by the sum of penalties for unsatisfied constraints, so that global minima correspond to the satisfaction of all constraints. For example, unjammed sphere packings are global minima of an energy that penalizes overlaps between spheres~\cite{ohern2003jamming,liu2010jamming}, self-assembled structures are global minima of a free energy composed of competing energies and entropies~\cite{murugan2014multifarious,weishun2017associative}, functional networks are global minima of cost functions that penalize the lack of function~\cite{rocks2018limits,murugan2018bioinspired}, while neural networks that correctly categorize data are global minima of cost functions that penalize incorrect identifications~\cite{choromanska2015loss}.

In all of these problems, the set of solutions (flat regions of the landscape corresponding to global minima) shrinks and divides as more constraints are added, until it completely disappears at the SAT/UNSAT transition \cite{kzrakala2007gibbs}. In the ideal case of an optimal algorithm, one always finds a solution up to the SAT/UNSAT transition. In actual practice, however, algorithms searching for solutions often encounter a transition below the SAT/UNSAT transition in which the problem changes from being easy to being hard to solve. The landscape becomes rough (see, e.g. \cite{wales2003energy,stillinger2016energy}) with many local minima. One approach in the "hard" phase is to choose an algorithm that can better avoid being trapped in local minima~\cite{torquato2000is,chaudhuri2010jamming,berthier2016equilibrium,mishkin2016all}.
Here we explore an alternate approach, where we transform the landscape to decrease the number of local minima without altering regions corresponding to global minima, to make the problem easier to solve for {\em any} practical algorithm.



We 
draw inspiration
from protein folding~\cite{onuchic1997theory}, where a specific protein evolves from an initial (denatured) configuration to the functional (native) state. Levinthal first noted \cite{levinthal1969fold} that an extensive entropy of local minima (undesired configurations) would prevent the protein from finding its native state in reasonable time. However, nature provides many proteins with cost functions (free energies) with landscapes that are partially smoothed out and tilted as a funnel towards the native state~\cite{onuchic1997theory,bryngelson1995funnels}.


In constructing cost functions, the penalties for unsatisfied constraints are usually equivalent. For example, in the jamming problem the pairwise energy costs for overlaps of spheres are identical.  The protein-folding problem suggests that by imposing non-equivalent penalties on unsatisfied constraints we can construct a cost function with a funneled landscape, in which the basin of attraction of global minima is increased at the cost of the existence or size of basins of local minima. In general, it is not obvious how the topography of the global energy landscape is modified by changes in local interactions \cite{wales2001microscopic,ruiz-garcia2017bifurcation}. Here we study two systems with rough landscapes (Fig. \ref{fig1}), namely tuned flow networks~\cite{rocks2018limits} and sphere packings~\cite{ohern2003jamming,liu2010jamming,goodrich2016scaling}. 
In both systems, we show that we can smooth out local minima, increasing the ability to reach global minima. An essential element of our approach is that we modify interactions in such a way as to leave global minima unaltered. As a result, it is not necessary to map back to the unweighted landscape, in contrast to previous approaches for smoothing energy landscapes~\cite{wales1999global,stillinger1988nonlinear,wawak1998diffusion}.


\section{Tuning of flow networks}

{\it A Flow network} is a set of nodes, each with a scalar pressure, and links (edges) between them that carry currents. The current through each edge is given by the product of the conductance and the pressure difference between the two nodes connected by the edge. Here we assume that all edges have the same conductance.
We refer the reader to the supplementary materials (SM) \cite{SM} for details of the ensemble of networks studied and the calculation of the pressure field. 
For a given initial network (\emph{e.g.} the one in Fig.~\ref{fig1}(a)), we drive flow through the network via a source edge with a unit pressure drop. The initial network is tuned, by removing/reinserting edges, to have pressure drops $\{\Delta P_i\}$ such that the fractional pressure drop change $(\Delta P_i - \Delta P^{0}_i)/\Delta P^{0}_i$ is at least of magnitude $\eta$ across a set of randomly chosen target edges $\{i\}$; $\Delta P^{0}_i$ are the initial pressure drops. To tune the system, we define a cost function to measure how far the system is from performing the task~\cite{rocks2018limits}:

\begin{figure*}
\includegraphics[width= 0.7\linewidth]{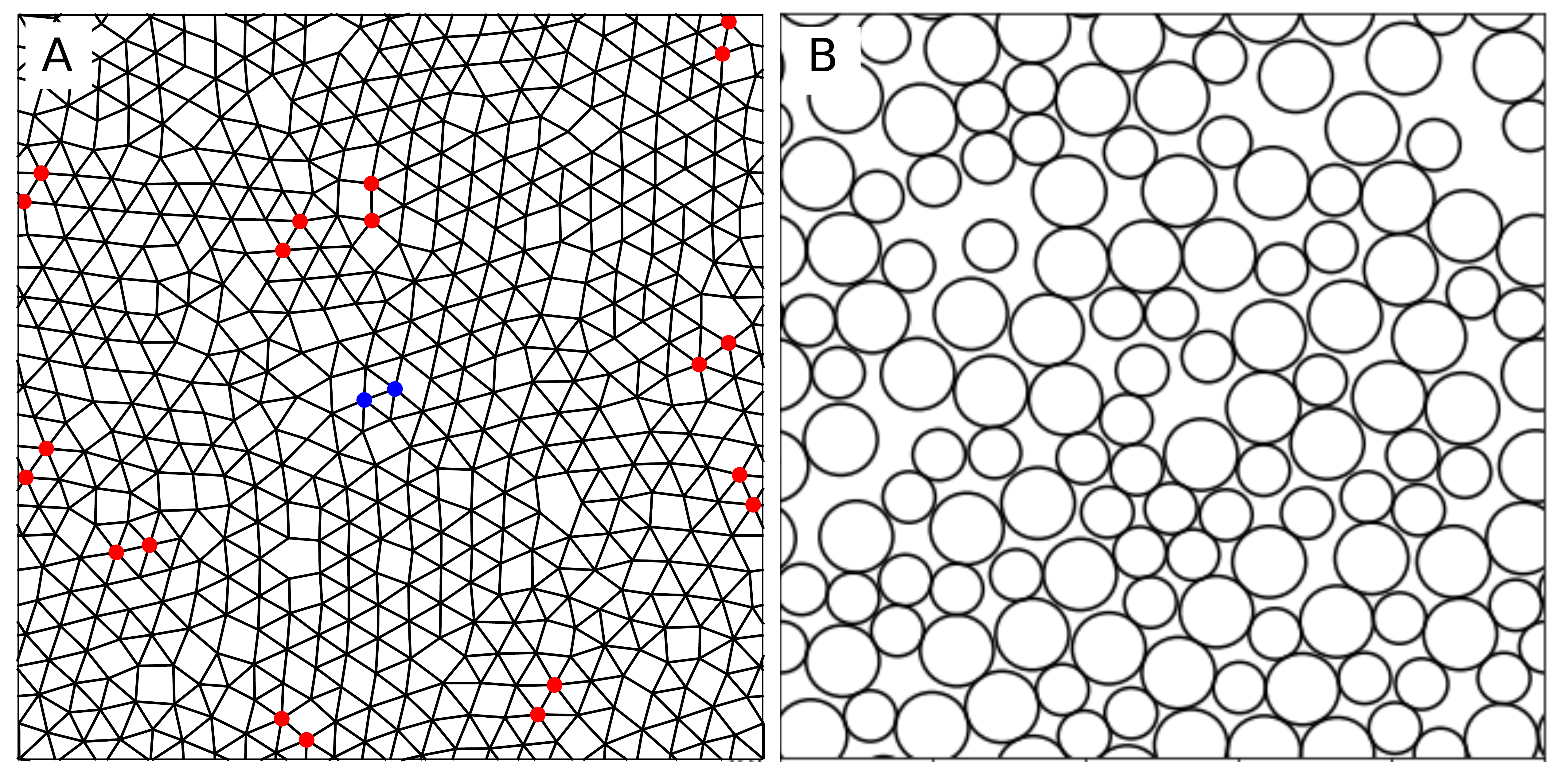}
\caption{ A) Tuning function into flow networks. Blue nodes in the center denote a ``source" edge with an externally-specified pressure difference; randomly-placed pairs of red nodes denote the ``target" edges, where we specify desired pressure drops. Edges may be removed or reinserted to change pressure drops across target edges. If all target pressure drops reach their desired values, the system reaches the global minimum of a cost function. B) Jamming of ideal spheres. Overlaps are minimized between bidisperse spheres with a diameter ratio of $1.4$ that are initially placed randomly. If all overlaps are eliminated, the system reaches the global minimum of the total energy and is unjammed.}
\label{fig1}
\end{figure*}

\begin{equation}
\mathcal{F} = \sum_{i=1}^{N_T}  r_i^2   \Theta(-r_i),
\label{eqn1}
\end{equation}
where $r_i=\frac{\Delta P_i - \Delta P^{0}_i}{\Delta P^{0}_i} - \eta$ is the deviation of the actual fractional change in the target pressure drop from the desired value $\eta$ for edge $i$. Here, $N_T$ is the number of target edges that we aim to tune for a system of $N$ nodes.  The Heaviside function ($\Theta$) in  $\mathcal{F}$ ensures that $\mathcal{F}=0$ if we have achieved at least a fractional change of pressure drop of $\eta$ for each target edge.  Since $\mathcal{F} \ge 0$,  $\mathcal{F}=0$ corresponds to the global minima of the cost function. This cost function has a complex landscape~\cite{Yan2017arquitecture}.

\begin{figure*}
\includegraphics[width= 0.7\linewidth]{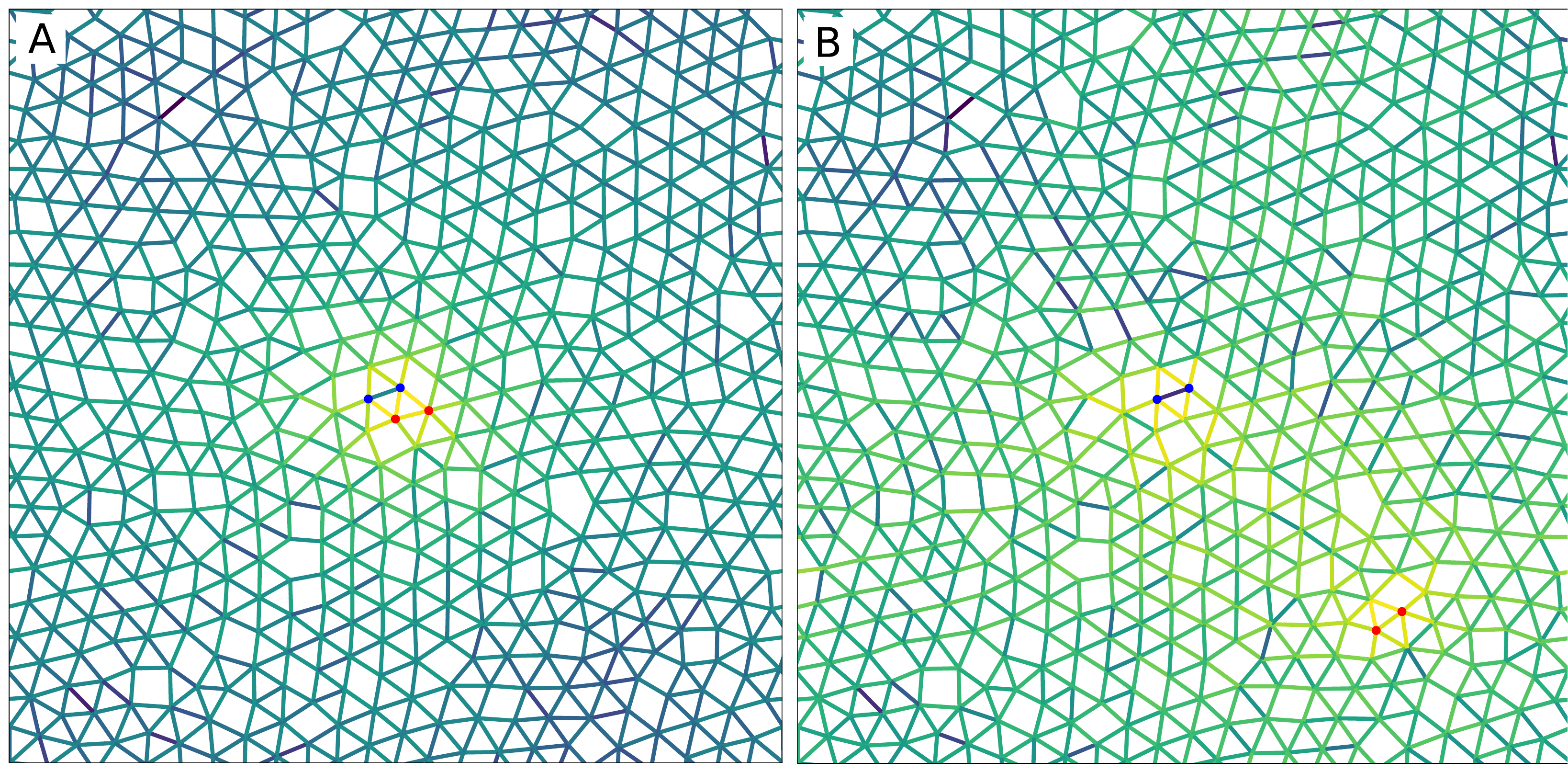}
\caption{Networks with one source edge and one target edge, with corresponding nodes labeled in blue and red, respectively. The color of each edge reflects the change in the pressure drop along the target edge if the given edge is removed. The target edge is close to the source in (A)  and far from it in (B).  See supplementary materials for an analytical approach to these cases \cite{SM}. }
\label{fig2}
\end{figure*}

The system is tuned using the greedy algorithm, by removing or reinserting edges from the initial network and always choosing the edge that reduces the value of $\mathcal{F}$ the most.  If  $\mathcal{F} =0$ the process is successful, and we say that the system can be tuned. If there are no bond deletions or reinsertions that would reduce the cost function and the value of $\mathcal{F}$ is greater than $0$, then the system is stuck in a local minimum and cannot be tuned successfully.

\begin{figure*}
\includegraphics[width= 0.75\linewidth]{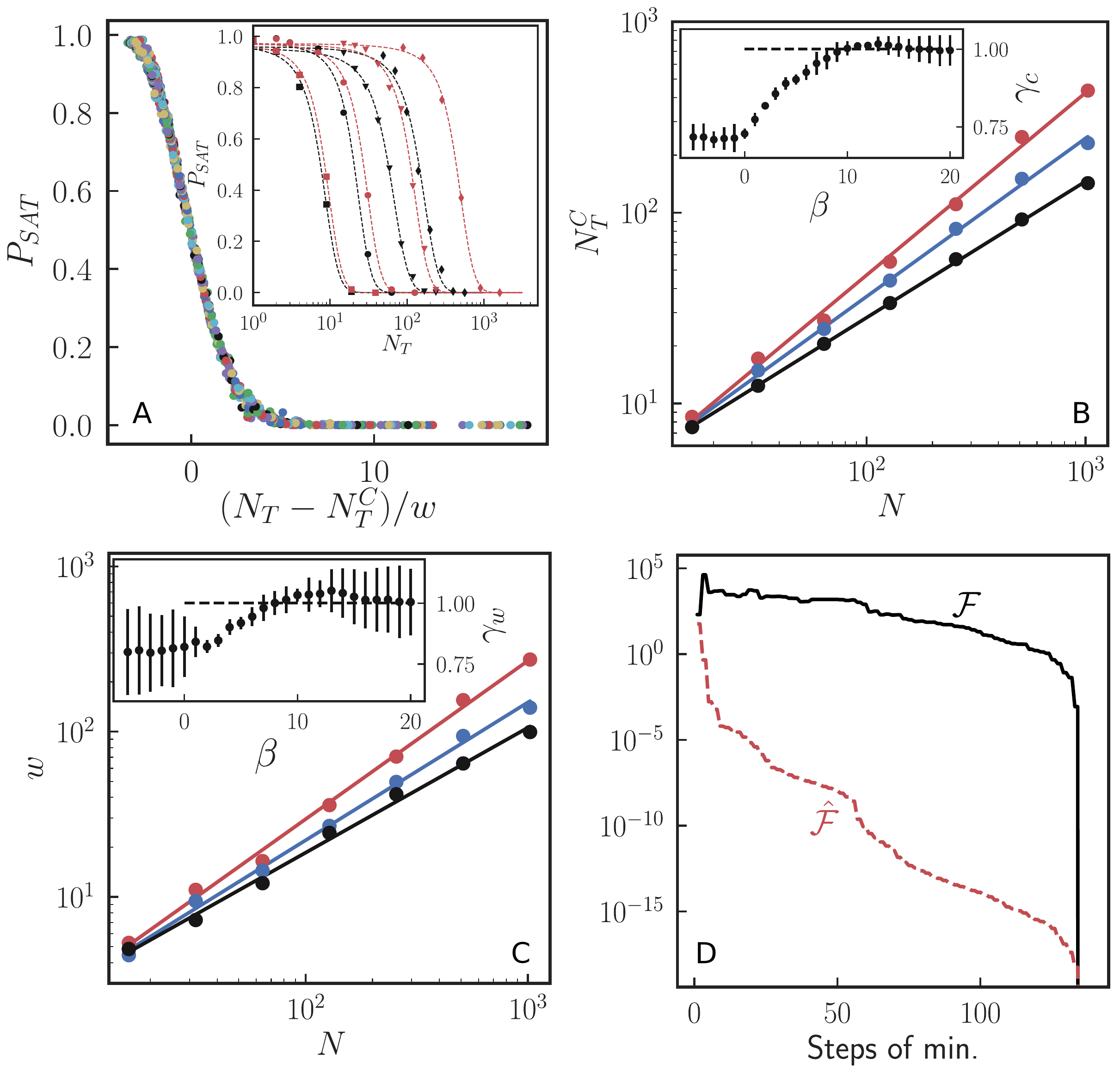}
\caption{Tuning flow networks in a funneled landscape. (A) Collapse of $P_{\text{SAT}}$ (the fraction of networks for which the system reaches the global minimum of $\mathcal{F}=0$) vs. the number of target edges, $N_T$, for $\beta=0, 2, 7$, $\eta = 0.1, 1$ and system sizes $N=16$, $32$, $64$, $128$, $256$, $512$ and  $ 1024$.  
Inset: $P_{\text{SAT}}$ vs. $N_T$ for $\eta = 0.1$ and $N=16$, $64$, $256$ and $1024$, labeled by squares, circles, triangles, and diamonds, respectively. Black and red curves correspond to $\beta=0$ and  $\beta=7$.
(B) Dependence on $N$ of the location of the 
transition, $N_T^c$ and (C) the transition width $w$ for $\eta=0.1$ and $\beta=0, 2, 7$ (black, blue and red points, respectively). Insets: Exponents $\gamma_c$ and $\gamma_w$ for the power-law fits of (B) and (C) plotted vs. $\beta$; error bars represent three times the standard deviation. (D) $\hat{\mathcal{F}}$ vs. minimization step as $\hat{\mathcal{F}}$ is minimized for $N=512$ nodes, $N_T=200$ targets, $\eta = 0.1$ and $\beta=7$ (red); we simultaneously calculate $\mathcal{F}$ ($\beta=0$) and show it for comparison (black). For analogous results for $\eta = 1$, see the SM~\cite{SM}.
}
\label{fig3}
\end{figure*}

In Ref.~\cite{rocks2018limits} it was shown that tuning of a complex flow or mechanical network exhibits a 
transition as the number of targets increases, where the problem becomes hard to solve.  Moreover, the maximum density of targets ($N_T/N$) that can be tuned successfully tends to zero as $N \to \infty$   (Fig. \ref{fig3}).  One would expect the maximum number of targets that can be tuned successfully to scale linearly in $N$. The observation of sublinear scaling cannot be explained by local geometrically frustrated motifs (such as having to tune three edges of a triangle), since the probability of such configurations for randomly chosen targets decreases with decreasing target density.  What is the source of frustration that prevents us from tuning the system in the thermodynamic limit?

Fig. \ref{fig2} suggests an answer. Here there is only one target edge (with target nodes labeled in red), and we color each edge by the magnitude of the pressure drop change \emph{at the target edge} if the colored edge is removed. In principle, every edge contributes, but not all edges contributed equally. Edges are colored on a blue to yellow scale where yellow edges give the largest changes in the target edge pressure drop $\Delta P$. Fig.~\ref{fig2} (A) shows that if the target edge is close to the source edge, only a few edges change $\Delta P$ significantly (only a few edges are yellow). By contrast, Fig.~\ref{fig2} (B) shows that if the target edge is far from the source edge, many different edges affect the $\Delta P$.  Evidently the source breaks translational symmetry for targets significantly. In particular, (1) if a target distant from the source is tuned first, subsequent tuning of a nearby target could significantly affect the distant target, causing failure in the tuning process; (2) If a nearby target is tuned first, there are still many edges available for tuning a distant target without affecting the nearby one, suggesting that the distant target can also be tuned successfully. These observations suggest that  tuning targets in order of their distance from the source could help. We transform the landscape using this information by modifying the cost function in Eq.~\eqref{eqn1} to 
 \begin{equation}
\hat{\mathcal{F} }= \sum_{i=1}^{N_T}  \frac{r_i^2}{R_i^\beta}   \Theta(-r_i),
 \label{eqn2}
 \end{equation}  
where $R_i$ is the distance of the $i$ target to the source and $\beta$ is an exponent that we can vary. For $\beta \gg 1$, the cost of incorrectly-tuned nearby targets is much higher than that of incorrectly-tuned faraway targets. Note that global minima in \eqref{eqn1} and \eqref{eqn2} are the same.

Results for different exponents $\beta$ in equation \eqref{eqn2} are shown in Fig. \ref{fig3}. Here $P_{\text{SAT}}$ is the fraction of networks that can be tuned; this is the success rate of reaching the global minima  ($\mathcal{F}=0$). We plot $P_{\text{SAT}}$ as a function of the number of target edges $N_T$. Fig. \ref{fig3} (A) shows that the curves collapse for all $\beta$, $\eta$ and system sizes $N$ studied, by introducing $N_T^c$, the number of targets that can be tuned when $P_\text{SAT}=0.5$, and the width $w$ of the $P_\text{SAT}$ curve corresponding to the spread in $N_T$ between $P_\text{SAT}=0.25$ and $P_\text{SAT}=0.75$. 
The inset to Fig. \ref{fig3}(A) shows that the $P_\text{SAT}$ curves shift to the right (more target edges can be tuned) as $\beta$ increases. As in Ref.~\cite{rocks2018limits}, we find power-law scaling: $N_T^c \sim N^{\gamma_c}$ and $w \sim N^{\gamma_w}$ with $\gamma_c \approx \gamma_w \approx 0.7$ for $\beta=0$.  As we increase $\beta$, $\gamma_c$ and $\gamma_w$ increase, saturating to unity (insets to Fig. \ref{fig3}B, C) so that $N_T^c/N \rightarrow \text{const}$ as $N \rightarrow \infty$. The scaling $N_T^c \sim N$ is consistent with tuning being limited only by local frustration that increases with target density but is independent of system size, suggesting that we have pushed the easy-to-hard transition up to the upper bound, the SAT/UNSAT transition, where the set of solutions disappears \cite{kzrakala2007gibbs}.  Finally, in Fig. \ref{fig3} (D) we plot the value of $\mathcal{F}$ (Eq.~\ref{eqn1}) and $\hat {\mathcal{F}}$ (Eq.~\ref{eqn2}) during the minimization of $\hat {\mathcal{F}}$ for one network. While $\hat{ \mathcal{F}}$ decreases monotonically to zero, $\mathcal{F}$ exhibits many local minima and energy barriers and stays approximately flat until it falls to zero precipitously, dropping more than $15$ orders of magnitude in the last few steps. This behavior demonstrates that we have indeed eliminated local minima, increasing the basin of attraction of global minima, showing that the landscape of $\hat{\mathcal{F}}$ is funneled. For more information see~\cite{SM}.

\vspace{1em}
\section{Jamming of ideal spheres}

{\it Jamming of ideal spheres} has been a useful starting point for studying disordered solids \cite{ohern2002random,ohern2003jamming,liu2010jamming}. We conduct numerical simulations on $50:50$ mixtures of spheres with a diameter ratio of $1.4$ in $d=2, 3$ spatial dimensions at fixed number density with periodic boundary conditions (see Fig. \ref{fig1}). In the standard procedure, one starts from $T=\infty$ with completely random particle positions and minimizes the total energy of the system: 
\begin{equation}
\mathcal{F} = \frac{1}{2 \alpha} \sum_{i \ne j}  \biggl(1 - \frac{|\vec{r}_i - \vec{r}_j|}{R_i+R_j}\biggr )^{\alpha} \Theta \biggl (1 - \frac{|\vec{r}_i - \vec{r}_j|}{R_i+R_j}  \biggr) ,
\end{equation}
where $\vec{r}_i$ is the position of the center of particle $i$ and $R_i$ is its radius. If $\mathcal{F}>0$ at the end of the minimization process, then the system is jammed and has not reached its global minimum, while if $\mathcal{F}=0$  (within a numerical tolerance) the system reaches its global minimum, an unjammed state.  We take $\alpha=2$, corresponding to harmonic repulsions between overlapping particles and use the FIRE algorithm \cite{bisek2006structural}. We are interested on the effect that a transformation of the landscape topography has over the properties of the jamming transition. We propose the new interparticle interaction:
 \begin{widetext}
 
\begin{align}
\hat{\mathcal{F}}=  &\frac{1}{2 \alpha} \sum_{i \ne j}                  \biggl\{ 1 + \beta \sum_{\substack{2D: s=x,y\\
                  3D: s=x,y,z}}
                  \left[  1 - \cos\left(\frac{2\pi}{L} s_{CM}\right) \right]\biggr\}
                  \biggl(1 - \frac{|\vec{r}_i - \vec{r}_j|}{R_i+R_j}\biggr )^{\alpha} \Theta \biggl (1 - \frac{|\vec{r}_i - \vec{r}_j|}{R_i+R_j}  \biggr),
                  \label{eqn4}
\end{align}
\end{widetext}
where $\vec{r}_{CM}=(\vec{r}_i+\vec{r'}_j)/2$ and $\vec{r'}_j$ is the periodical image of $\vec{r}_j$ closest to $\vec{r}_i$. The case $\beta=0$ corresponds to the original jamming landscape, and $\beta>0$ 
makes the interactions stronger at the center of the box and weaker at the corners
. As the energy $\hat {\mathcal{F}}$ is minimized, we expect particles to rearrange from the center outwards in order to eliminate overlaps, in contrast to the usual case of $\beta = 0$, where rearrangements occur everywhere in the system at the same time.

\begin{figure*}
\includegraphics[width= 0.75\linewidth]{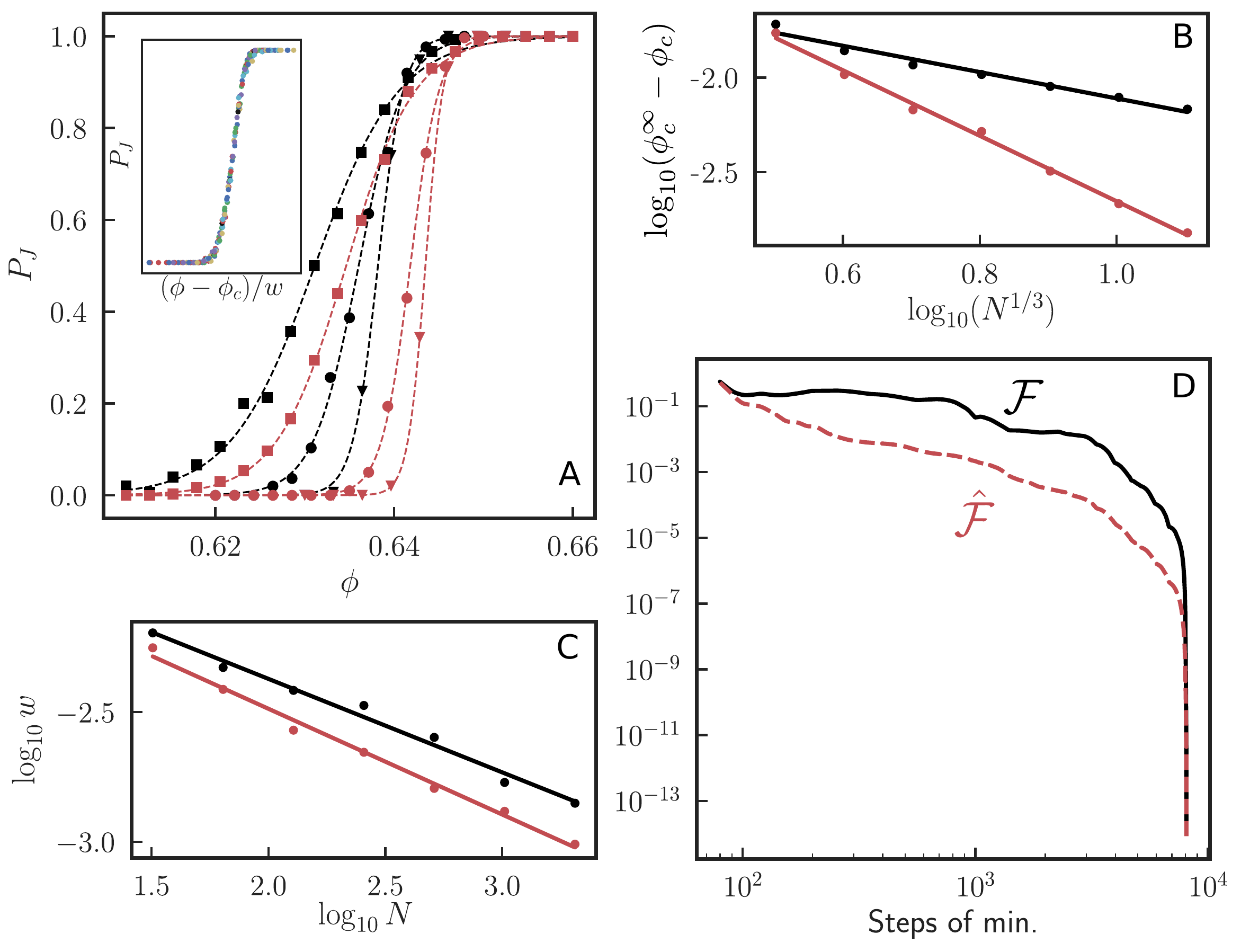}
\caption{Jamming of $d=3$ soft spheres in a smoother landscape. A) Probability of jamming, $P_J$, versus packing fraction $\phi$ for system sizes $N=64$ (squares), $512$ (circles) and $2048$ (triangles). Black and red lines correspond to $\beta=0$ and $\beta= 10$, respectively. 
Inset: Collapsed $P_J$ vs. $\phi$ curves for  $\beta=0, 10$ and sizes $32$, $64$, $128$, $256$, $512$, $1024$, and $2048$.
(B) Critical packing fraction (defined by $P_J (\phi_c) = 0.5$) vs. system size $N$. For $\phi_c^{\infty} = 0.645$, we find exponent values $-0.70\pm 0.05$ for $\beta=0$ and $-1.73 \pm 0.05$ for $\beta=10$. (C) Power law behavior for the width $w$ of the distribution vs. $N$; the straight lines have similar exponents ($-0.36 \pm0.02$ for $\beta=0$  and $-0.41 \pm 0.02$ for $\beta=10$). (D) Energy vs. minimization step for $N=32$ particles with $\phi = 0.63$ and  $\beta=30$. We use the modified potential energy $\hat{\mathcal{F}}$ (red) and also plot $\mathcal{F}$ along the minimization trajectory for comparison (black). 
}
\label{fig4}
\end{figure*}

Fig. \ref{fig4} shows results for $d=3$ case; for $d=2$ see the SM \cite{SM}. Fig.~\ref{fig4} (A) shows that the probability of jamming $P_J$ vs. packing fraction $\phi$ shifts to the right as $\beta$ is increased from $\beta=0$ (black) to $\beta=10$ (red) for each system size studied.  The curves for $P_J$ vs. $\phi$ can be collapsed~\cite{vagberg2011finite,graves} by introducing the position $\phi_c$ and the width $w$ of the jamming transition for each $N$.  Fig.~\ref{fig4} (B) shows $(\phi_c-\phi_c^\infty)$ vs. $N$; here, we assume that the position of the jamming transition in the thermodynamic limit, $\phi_c^\infty$, is unaffected by $\beta$. (Alternatively, we could assume that $\phi_c^\infty$ depends on $\beta$; that analysis is shown in the SM~\cite{SM}.) The scaling steepens with $\beta$ and the prefactor of the scaling of the transition width decreases, suggesting this is a smoother landscape. Fig.~\ref{fig4} (D) shows this explicitly; during one minimization, we plot $\hat {\mathcal{F}}$, the quantity that we actually minimize,  and simultaneously show $\mathcal{F}$. As before, $\hat {\mathcal{F}}$ decreases monotonically while $\mathcal{F}$ exhibits energy barriers. 


Our results show that equation \eqref{eqn4} reduces the ratio of the volume occupied by basins of local minima compared to the volume corresponding to global minima. This shows that the landscape is funneled. It would be interesting to quantify the structure of the new landscape~\cite{wales2003energy,stillinger2016energy}, possibly by measuring the distribution of basin volumes of local minima~\cite{xu2011direct,martinianifrenkel}.
While the shifts in the critical packing fraction are not large compared to those achieved by Monte Carlo swap methods~\cite{ozawa2017exploring}, indicating that the implementation of the funnel is less effective than for the tuning problem, our results nevertheless show a significant increase in the probability of reaching an unjammed state ($1-P_J$) at a given packing fraction $\phi$ at a given system size $N$.

\section{Discussion}

In this paper we have transformed the landscapes of two completely different systems, smoothing out local minima and increasing the basins of attraction of global minima. This conserves global minima but modifies the topography of the landscape away from them, thus shifting the transition where the problem becomes hard to solve.  We emphasize that in both cases, the transition from easy- to hard-to-solve depends not only on the structure of global minima, corresponding to zeroes of the cost function~\cite{kzrakala2007gibbs}, but also on nonzero values of the cost function (particularly local minima). In flow networks, all the constraints involve all of the edges of the network, but for constraints near the source, fewer edges contribute heavily. Thus, the source breaks translational symmetry and provides a natural geometrical choice for the constraint weightings. In jamming, on the other hand, the system is isotropic on average and each constraint involves only one pair of particles, so it is less clear how to choose a useful weighting of constraints. We break this invariance arbitrarily by introducing a transformation that picks out a region in actual physical space to minimize first. This smooths out the landscape, but not to the same degree as for flow networks.

We can think about the tuning of flow networks in a different way, in terms of how the satisfaction of each constraint affects satisfiability of other constraints. If the degrees of freedom that contribute heavily to a constraint also contribute heavily to a different constraint, we say those two constraints are correlated with each other--the satisfaction of one affects the ability to satisfy the other. This correlation is directional if the satisfaction of one constraint affects the ability to satisfy another one more than in the reverse case (as in Fig. \ref{fig2}). Flow networks teach us that it is useful to weight constraints that share heavily-contributing degrees of freedom together, and that a broken symmetry (like the one introduced by the source) can indicate a preferred order between these groups. More generally, however, it suggests a strategy of weighting together constraints that are highly correlated with each other, even in cases where there is no preferred order. This strategy can be used for jamming or other cases in which there is parity, or approximate parity, in how many degrees of freedom contribute heavily to the constraints.

This insight is potentially generalizable to other processes driven by landscape optimization, such as self-assembly~\cite{murugan2014multifarious,weishun2017associative}, machine learning~\cite{choromanska2015loss,geiger2018jamming}, discrete constraint-satisfaction problems in computer science~\cite{mezard2009information} or signal reconstruction ~\cite{krzakkala2012statistical}. The introduction of funnels, or other transformations to cost functions, may lead to more effective ways of reaching global minima and to a better understanding of the role played by landscape topography when tackling constraint satisfaction problems.

We thank H.-H. Boltz and J. W. Rocks for helpful discussions, and to H. Ronellenfitsch and H.-H. Boltz for providing initial versions of the code. This research was supported by the the National Science Foundation via DMR-1506625 (MRG), the Burroughs-Wellcome Fund (MRG,EK), and the Simons Foundation via awards 327939 (AJL) and 454945 (MRG, AJL).


\begin{thebibliography}{99}

\bibitem{liu2010jamming}
A. J. Liu and S. R. Nagel, 
 \href{https://www.annualreviews.org/doi/abs/10.1146/annurev-conmatphys-070909-104045}{Annual Review of Condensed Matter Physics, {\bf 1}, 347 (2010) }
 
\bibitem{ohern2003jamming} 
C. S. O'Hern, L. E. Silbert, A. J. Liu, and S. R. Nagel, 
\href{https://journals.aps.org/pre/abstract/10.1103/PhysRevE.68.011306}{Phys. Rev. E {\bf 68}, 011306 (2003) }


\bibitem{weishun2017associative}
W. Zhong, D. J. Schwab and A. Murugan, 
\href{https://doi.org/10.1007/s10955-017-1774-2}{J. Stat. Phys., {\bf 167}, 806 (2017)}

\bibitem{murugan2014multifarious}
A. Murugan, Z. Zeravcic, M. P. Brenner and S. Leibler, 
\href{http://www.pnas.org/content/112/1/54}{Proc. Natl. Acad. Sci. USA, {\bf 112}, 54 (2015)}


\bibitem{rocks2018limits} 
J. W. Rocks, H. Ronellenfitsch, A. J. Liu, S. R. Nagel and E. Katifori,
\href{https://www.pnas.org/content/116/7/2506}{Proc Natl Acad Sci, \textbf{116}, 2506 (2019)}

\bibitem{murugan2018bioinspired}
A, Murugan, H. M. Jaeger,
\href{https://www.cambridge.org/core/journals/mrs-bulletin/article/bioinspired-nonequilibrium-search-for-novel-materials/3058E8856513C99E44014438FCF76231}{MRS Bulletin {\bf 44}, 96 (2018) }


%


\bibitem{choromanska2015loss}
A. Choromanska, M. Henaff, M. Mathieu, G. B. Arous, and Y. LeCun,  
\href{http://proceedings.mlr.press/v38/choromanska15.pdf}{In Artificial Intelligence and Statistics, 192, (2015)}


\bibitem{kzrakala2007gibbs}
F. Krzakala, A. Montanari, F. Ricci-Tersenghie, G. Semerjianc, and L. Zdeborov\'{a}f,  
\href{https://www.pnas.org/content/104/25/10318.short}{Proc Natl Acad Sci, \textbf{104}, 10318 (2007)}



\bibitem{wales2003energy}
D. J. Wales, 
{\it Energy landscapes},
 \href{https://www.cambridge.org/core/books/energy-landscapes/8389839B8AA8E4ADCF0281C37481FF3B}{Cambridge University Press (2003)}



\bibitem{stillinger2016energy}
F. H. Stillinger,
 {\it Energy landscapes, inherent structures, and condensed-matter phenomena}, \href{https://press.princeton.edu/titles/10643.html}{Princeton University Press (2016)}
 
\bibitem{berthier2016equilibrium}
L. Berthier, D. Coslovich, A. Ninarello, and M. Ozawa,
\href{
https://journals.aps.org/prl/pdf/10.1103/PhysRevLett.116.238002}{Phys. Rev. Lett., \textbf{116}, 238002 (2016)}



\bibitem{mishkin2016all}
D. Mishkin, J. Matas, 
\href{https://arxiv.org/pdf/1511.06422.pdf}{ArXiv: 1511.06422 (2016)}



\bibitem{torquato2000is}
S. Torquato, T. M. Truskett, and P. G. Debenedetti, 
\href{https://journals.aps.org/prl/abstract/10.1103/PhysRevLett.84.2064}{Phys. Rev. Lett., \textbf{84}, 2064 (2000)}



\bibitem{chaudhuri2010jamming}
P. Chaudhuri, L. Berthier, and S. Sastry,
\href{https://journals.aps.org/prl/pdf/10.1103/PhysRevLett.104.165701}{Phys. Rev. Lett., \textbf{104}, 165701 (2010)}

 
\bibitem{onuchic1997theory}
J. N. Onuchic, Z. Luthey-Schulten and P. G. Wolynes, 
 \href{https://www.annualreviews.org/doi/abs/10.1146/annurev.physchem.48.1.545}{ Annu. Rev. Phys. Chem. {\bf 48}, 545  (1997) }


 \bibitem{levinthal1969fold}
 C. Levinthal, {\it How to fold graciously}. 
P. De- Brunner, J. Tsibris, and E. Munck, (eds.). Urbana, IL University of Illinois Press, 22-24 (1969)
 
\bibitem{bryngelson1995funnels}
J. D. Bryngelson, J. N. Onuchic, N. D. Socci, and P. G. Wolynes, 
 \href{https://onlinelibrary.wiley.com/doi/abs/10.1002/prot.340210302}{ PROTEINS: Structure, Function, and Genetics, {\bf 21}, 167  (1995) }


\bibitem{wales2001microscopic}
D. J. Wales, 
 \href{http://science.sciencemag.org/content/293/5537/2067}{ Science, {\bf 293}  (2001) }

\bibitem{ruiz-garcia2017bifurcation}
M. Ruiz-Garcia, L. L. Bonilla and  A. Prados, 
\href{https://journals.aps.org/pre/abstract/10.1103/PhysRevE.96.062147}{Phys. Rev. E {\bf 96}, 062147 (2017) }





\bibitem{goodrich2016scaling} 
C. P. Goodrich, A. J. Liu and J. P. Sethna,
\href{https://www.pnas.org/content/pnas/113/35/9745.full.pdf}{Proc Natl Acad Sci, {\bf 113}, 9745 (2016) }.


\bibitem{wales1999global}
D. J. Wales and H. A. Scheraga,
\href{http://science.sciencemag.org/content/285/5432/1368}{Science, {\bf 285}, 1368 (1999)}

\bibitem{stillinger1988nonlinear}
F. H. Stillinger and T. A.  Weber, 
\href{ https://doi.org/10.1007/BF01011658}{J. Stat. Phys., {\bf 52}, 1429 (1988)}

\bibitem{wawak1998diffusion}
R. J. Wawak, J. Pillardy, A. Liwo, K. D. Gibson and
H. A. Scheraga, 
\href{https://pubs.acs.org/doi/abs/10.1021/jp972424u}{J. Phys. Chem. A, {\bf 102}, 2904 (1998)}



\bibitem{SM}  See Supplemental Material at [URL will be inserted by publisher]. It contains details about the systems used in the simulations, analogous simulations to the ones shown in the main text but for different sets of parameters and analytical arguments related to Fig. \ref{fig2}. Finally, we have also included three videos showing the tuning of flow networks.


\bibitem{Yan2017arquitecture}
L. Yan, R. Ravasio, C. Brito, and M. Wyart, 
\href{https://www.pnas.org/content/114/10/2526.short}{Proc. Natl. Acad. Sci. USA, 114(10), 2526 (2017)}



\bibitem{ohern2002random}
C. S. O'Hern, S. A. Langer, A. J. Liu and S. R. Nagel,
\href{https://journals.aps.org/prl/abstract/10.1103/PhysRevLett.88.075507}{Phys. Rev. Lett., {\bf 88}, 075507 (2002)}



\bibitem{bisek2006structural} 
E. Bitzek, P. Koskinen, F. G\"{a}hler, M. Moseler, and P. Gumbsch, 
\href{https://journals.aps.org/prl/abstract/10.1103/PhysRevLett.97.170201}{Phys. Rev. Lett. {\bf 97}, 170201 (2006) }
 
 
\bibitem{graves}
A. L. Graves, S. Nashed, E. Padgett, C. P. Goodrich, A. J. Liu, J. P. Sethna, 
\href{https://journals.aps.org/prl/abstract/10.1103/PhysRevLett.116.235501}{Phys. Rev. Lett., {\bf 116}, 235501 (2016)}


 \bibitem{vagberg2011finite} 
D. V\aa gberg, D. Valdez-Balderas, M. A. Moore, P. Olsson, and S. Teitel, 
\href{https://journals.aps.org/pre/abstract/10.1103/PhysRevE.83.030303}{Phys. Rev. E  {\bf 83}, 030303(R) (2011) }





 
\bibitem{xu2011direct}
N. Xu, D. Frenkel and A. J. Liu,
\href{https://journals.aps.org/prl/abstract/10.1103/PhysRevLett.106.245502}{Phys. Rev. Lett. {\bf 106}, 245502 (2011) }

\bibitem{martinianifrenkel}
S. Martiniani, K. J. Schrenk, J. D.  Stevenson, D. J. Wales, D. Frenkel, 
\href{https://journals.aps.org/pre/abstract/10.1103/PhysRevE.93.012906}{Phys. Rev. E, {\bf 93}, 012906 (2016)}




%
\bibitem{ozawa2017exploring} 
M. Ozawa, L. Berthier and D. Coslovich,
\href{https://scipost.org/SciPostPhys.3.4.027/pdf}{SciPost Phys.  {\bf 3}, 027 (2017) }


\bibitem{geiger2018jamming}
M. Geiger, S. Spigler, S. d'Ascoli, L. Sagun, M. Baity-Jesi, G. Biroli and M. Wyart, 
\href{https://arxiv.org/pdf/1809.09349.pdf}{ArXiv: 1809.09349 (2018)}

\bibitem{mezard2009information}
M. Mezard and A. Montanari, 
{\it Information, Physics and Computation},
 \href{https://global.oup.com/academic/product/information-physics-and-computation-9780198570837?cc=us&lang=en&}{Oxford Graduate Texts (2003)}

\bibitem{krzakkala2012statistical}
F. Krzakala, M. M\'ezard, F. Sausset, Y. F. Sun and L. Zdeborov\'a, 
\href{https://journals.aps.org/prx/abstract/10.1103/PhysRevX.2.021005}{Phys. Rev. X, {\bf 2}, 021005 (2012)}


 






%
%

%
 
 



\end{thebibliography}
\end{document}